\documentclass[showpacs,preprintnumbers,preprint,nofootinbib]{revtex4}
\usepackage{graphicx}
\usepackage[normalem]{ulem}
\begin{document}
\title{Complex scaling method for three- and four-body scattering above the break-up thresholds}
\author{Rimantas Lazauskas}
\email{rimantas.lazauskas@iphc.cnrs.fr}
\affiliation{IPHC, IN2P3-CNRS/Universit\'e Louis Pasteur BP 28, F-67037 Strasbourg Cedex
2, France}
\author{Jaume Carbonell}
\email{jaume.carbonell@ipno.in2p3.fr}
\affiliation{Institut de Physique Nucl\'eaire Orsay, CNRS/IN2P3,
F-91406 Orsay Cedex, France.}
\date{\today}

\begin{abstract}
A formalism based on the complex-scaling method is presented to solve the few particle
scattering problem in configuration space using bound state techniques with trivial
boundary conditions.
Several applications to A=3,4 systems are presented to demonstrate the efficiency of the method
in computing elastic as well as break-up reactions with Hamiltonians including both short and long-range interaction.
\end{abstract}

\maketitle

\section{Introduction}
\label{intro}

The theoretical description of the quantum-mechanical collisions is one of the most complex problems in theoretical physics.
The main difficulty to solve the scattering problem in configuration space is related to the fact that, unlike the bound state
case, the scattering wave functions are not localized. One is therefore obliged to solve multidimensional integro-differential
equations with extremely complex boundary conditions. Therefore, finding a method which could enable us to solve the scattering
problem without an explicit use of the asymptotic form of the wave function is of great importance.  Recently, the research in this
field has been intensified very much~\cite{Barnea:2001prc,Uzu_prc68:2003,Rubtsova_prc79:2009,Navratil_prc82:2010,Laza_csm:2011,Pisa_prc85:2012,Arnas_cem:2012}.

One of the pioneering works in this direction is the development of the Lorentz integral  transform~\cite{Barnea:2001prc},
which allows to calculate the integral cross section of a scattering process using bound state like techniques.
Still its application  to compute differential observables becomes rather involved.
Few years later, the complex energy method has been applied to the four-nucleon problem with separable potential~\cite{Uzu_prc68:2003},
providing full information on the scattering process. Lately, after some technical improvements,
this method was succesfully applied   to describe realistic four-nucleon break-up process~\cite{Arnas_cem:2012}.
Other recent developments include momentum lattice technique~\cite{Rubtsova_prc79:2009} and a method based on
discretized continuum solutions~\cite{Pisa_prc85:2012},   although they are no yet tested in the four-body sector and to the break-up case respectively.

On the other hand, already in the late sixties, Nuttal and Cohen~\cite{Nuttal_csm} proposed a very handy method to
treat the scattering problem for short range potentials using coordinate space variational bound state techniques, namely
the complex scaling method. However application of this method to the scattering problem has lingered.
Only recently a variant of the complex scaling method to calculate scattering observables above the break-up threshold has been
proposed and applied by Giraud et al.~\cite{GKO_04}. Still this variant relies on the spectral function formalism and
requires a diagonalization of the full N-body matrices to get converged results,
which makes it difficult to extend beyond the N=3 case.

In~\cite{Laza_csm:2011} we have demonstrated that only  a slightly modified original version of the complex scaling method
by Nuttal and Cohen~\cite{Nuttal_csm}, if combined with Fadddeev equations, turns to be very efficient
in describing three-nucleon scattering, including break-up reactions.
We have shown that this method may treat strong interaction of almost any complexity:
realistic local and non-local potentials,
optical potentials, all of them in conjunction with repulsive Coulomb force.
The method allows to calculate both
elastic as well as break-up amplitudes, thus providing a full description of the three-body collisions.
There are no formal obstacles in extending this method to treat collisions involving any number of particles,
as long as one is able to handle the eventual large scale numerical problem.
In this contribution we will summarize this current effort and present some new results concerning the solution of the 4-nucleon
scattering problem above the breakup thresholds.

\section{Formalism}

\begin{figure}[h!]
\begin{center}
\includegraphics[width=0.75\textwidth]{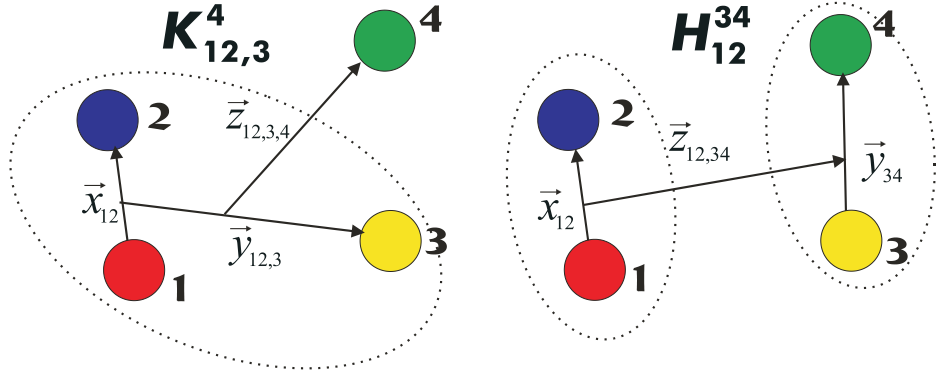}
\end{center}
\caption{
The FY components $K_{12,3}^4$ and $H_{12}^{34}$ for a given particle ordering. As $z\rightarrow \infty$,
the $K$ components describe 3+1 particle channels, while the $H$ components contain asymptotic states of 2+2 channels.}
\label{Fig_4b_config}
\end{figure}

In our previous paper~\cite{Laza_csm:2011} we have presented the basic ideas of the complex scaling method used to solve the scattering
problem for 2- and 3-particle systems. Here, without loosing generality, we will briefly summarize the formalism
 for solving   the Faddeev-Yakubovski (FY) equations  for a system of four identical particles.
In this particular case, we start by separating the incoming plane wave from the FY components (see figure~\ref{Fig_4b_config}):
\begin{eqnarray}\label{In_wave_sep}
K(\vec{x},\vec{y},\vec{z})=K^{out}(\vec{x},\vec{y},\vec{z})+K^{in}(\vec{x},\vec{y},\vec{z}) \\
H(\vec{x},\vec{y},\vec{z})=H^{out}(\vec{x},\vec{y},\vec{z})+H^{in}(\vec{x},\vec{y},\vec{z}).
\end{eqnarray}
Note, that the four-body FY equations involve components of two types, K and H,
in the asymptote describing the elastic 3+1 and 2+2 particle channels respectively.
Therefore, in the last equation, one has $H^{in}\equiv0$  by considering collision of 3+1 particle
clusters  whereas $K^{in}\equiv0$ by considering 2+2 particle collisions. Asymptotes of the FY components
$K^{out}$ and $H^{out}$ will contain only various combinations of the outgoing waves.
These components are solutions of the driven FY equations, which have the form:
\begin{eqnarray}
\nonumber \left( E-H_{0}-V_{12}\right) K^{out}&-&V_{12}(P^{+}+P^{-})\left[ (1+Q)K^{out}+H^{out}\right]\\
&&=V_{12}(P^{+}+P^{-})\left[ (1+Q)H^{in}+QK^{in}\right], \label{FY_drive1} \\
\left( E-H_{0}-V_{12}\right) H^{out}&-&V_{12}\tilde{P}\left[ (1+Q)K^{out}+H^{out}\right]
=V_{12}\tilde{P}\left[ (1+Q)K^{in}\right] ,
\label{FY_drive2}
\end{eqnarray}

It is interesting to introduce the complex scaling operator (CSO)
\begin{equation}
\widehat{S}=e^{i\theta r\frac{\partial }{\partial r}}=e^{i\theta(x\frac{\partial }{\partial x}+y\frac{\partial }{\partial y}+z\frac{\partial }{\partial z})},  \label{CSO}
\end{equation}
where $r^2=x^2+y^2+z^2$ and $\theta$ is the complex scaling angle, of free choice. The action of the complex scaling
operator on outgoing wave gives
\begin{equation}
\widehat{S}\exp(ikr)=\exp(-kr\sin{\theta})\exp(ikr\cos{\theta}).
\label{CSO_action}
\end{equation}
Therefore if one acts on $K^{out}$ (or $H^{out}$)
with the complex scaling operator (\ref{CSO}), by keeping the
complex scaling angle in the range $0<\theta<\pi/2$, one
gets exponentially decreasing functions $\widetilde{K^{out}}\equiv\widehat{S}K^{out}$ and  $\widetilde{H^{out}}\equiv\widehat{S}H^{out}$.
By acting with the same operator on equations~(\ref{FY_drive1}-\ref{FY_drive2}),  one gets equivalent equations  for the
scaled FY components $\widetilde{K^{out}}$ and $\widetilde{H^{out}}$\footnote{Note, that the four-body incoming wave components $K^{in}$
(or $\widetilde{K^{in}}$) and $H^{in}$ (or $\widetilde{H^{in}}$) are constructed from the corresponding bound-state wave functions of  3- and 2-particle
systems respectively.}:
\begin{eqnarray}
\nonumber \widehat{S}\left( E-H_{0}-V_{12}\right)\widehat{S}^{-1} \widetilde{K^{out}}-\widehat{S}V_{12}\widehat{S}^{-1} (P^{+}+P^{-})\left[ (1+Q)\widetilde{K^{out}}+\widetilde{H^{out}}\right] \\
=\widehat{S}V_{12}\widehat{S}^{-1} (P^{+}+P^{-})\left[ (1+Q)\widetilde{H^{in}}+Q\widetilde{K^{in}}\right],\label{FY_drive_cs1} \\
\nonumber \widehat{S}\left( E-H_{0}-V_{12}\right)\widehat{S}^{-1}  \widetilde{H^{out}}-\widehat{S}V_{12}\widehat{S}^{-1} \tilde{P}\left[ (1+Q)\widetilde{K^{out}}+\widetilde{H^{out}}\right] \\
=\widehat{S}V_{12}\widehat{S}^{-1} \tilde{P}\left[ (1+Q)\widetilde{K^{in}}\right] ,
\label{FY_drive_cs2}
\end{eqnarray}
The functions to determine in the last equation, notably $\widetilde{K^{out}}$ and $\widetilde{H^{out}}$, are exponentially bound. Thus,
in principle, bound-state  techniques based on square integrable basis functions, may be employed to solve these equations.
Nevertheless, one should still ensure that the
inhomogeneous terms on the right hand side of eqs.~(\ref{FY_drive_cs1}-\ref{FY_drive_cs2}) is also square integrable. Indeed,  after the
complex scaling, the incoming waves $\widetilde{K^{in}}$ and $\widetilde{H^{in}}$, diverge when the intercluster separation increases. This divergence
should be screened by the short range potential term $\widehat{S}V_{12}\widehat{S}^{-1}$. Nevertheless, the direction in which propagates
the incoming wave does not coincide completely with the direction of the potential term.  For a system of four identical particles,
the inhomogeneous terms are exponentially bound if the following conditions are satisfied:
\begin{equation}
\tan{\theta}<\sqrt{\frac{2B_3}{E_{cm}}},
\label{lim_31}
\end{equation}
in case of incoming wave of $1+3$-type --  $B_3$ being the binding energy of 3-particle cluster -- and
\begin{equation}
\tan{\theta}<\sqrt{\frac{2B_2}{E_{cm}}},
\label{lim_22}
\end{equation}
when considering incoming wave of $2+2$-type with a binding energy of 2-particle cluster equal $B_2$.
$E_{cm}$ denotes the energy of the two scattering clusters in the center of mass frame. Similar conditions for a 3-body
problem have been derived in~\cite{Laza_csm:2011}.

The scattering amplitudes are extracted from the solutions of  eqs.~(\ref{FY_drive_cs1}-\ref{FY_drive_cs2})
by using integral relations, obtained through the Green's theorem~\cite{Laza_csm:2011,Laza_csm:2012}.
Equations~(\ref{FY_drive_cs1}-\ref{FY_drive_cs2}) are solved using the method described in our previous
works~\cite{These_Rimas_03,Lazauskas_4B,Laza_csm:2011}. Namely, first spin, isospin and angular dependence of
the FY components is expanded using partial-waves. The radial dependence of the FY components, in variables (x,y,z),
is expanded by means of cubic spline basis and vanishing boundary conditions are implemented at the borders of the chosen
3-dimensional grid.

\section{Results}

We will present in this section some chosen results obtained by applying the complex scaling method to 3- and 4-body systems.
In table~\ref{tab:ndpd} we summarize the phase shifts and inelasticity parameters calculated for S-wave nucleon-deuteron scattering, using MT I-III
potential~\cite{MT13}. Very nice agreement is obtained compared to the  benchmark calculation of reference~\cite{Arnas_bench,Friar_nd_bench}, both by neglecting (n-d)
as well as including (p-d) repulsive Coulomb interaction. Differential break-up amplitudes may be equally extracted
with very nice accuracy~\cite{Laza_csm:2011}, although they are not reported in this work.
\begin{table}
\caption{Phase shifts and inelasticity parameters calculated for S-wave nucleon-deuteron scattering, using MT I-III
potential~\cite{MT13}}\label{tab:ndpd}
\begin{tabular}{lllllll}
\hline\noalign{\smallskip}
  & & & \multicolumn{2}{c}{n-d} & \multicolumn{2}{c}{p-d}  \\
P.W. & $E_{lab}$ (MeV)& & This work & Ref.~\cite{Arnas_bench,Friar_nd_bench} & This work & Ref.~\cite{Arnas_bench} \\
\noalign{\smallskip}\hline\noalign{\smallskip}

$^2S_{\frac{1}{2}}$ & 14.1 & $Re(\delta)$ & 105.5 & 105.49 &108.4 &108.41[3]\\
                    &      & $\eta$ & 0.465 & 0.4649 & 0.498& 0.4983[1] \\
$^4S_{\frac{3}{2}}$ & 14.1 & $Re(\delta)$ & 69.0 & 68.95 &72.6 &72.60\\
                    &      & $\eta$ & 0.978 & 0.9782 & 0.983& 0.9795[1]\\
$^2S_{\frac{1}{2}}$ & 42.0 & $Re(\delta)$ & 41.5 & 41.35 &43.8 & 43.68[2]\\
                    &      & $\eta$ & 0.502 & 0.5022 &0.505 &0.5056\\
$^4S_{\frac{3}{2}}$ & 42.0 & $Re(\delta)$ & 37.7 & 37.71 &40.1 &39.96 [1]\\
                    &      & $\eta$ & 0.903 & 0.9033 & 0.904& 0.9046\\
\noalign{\smallskip}\hline
\end{tabular}
\end{table}

\begin{figure}
\begin{center}
\includegraphics[scale=0.47]{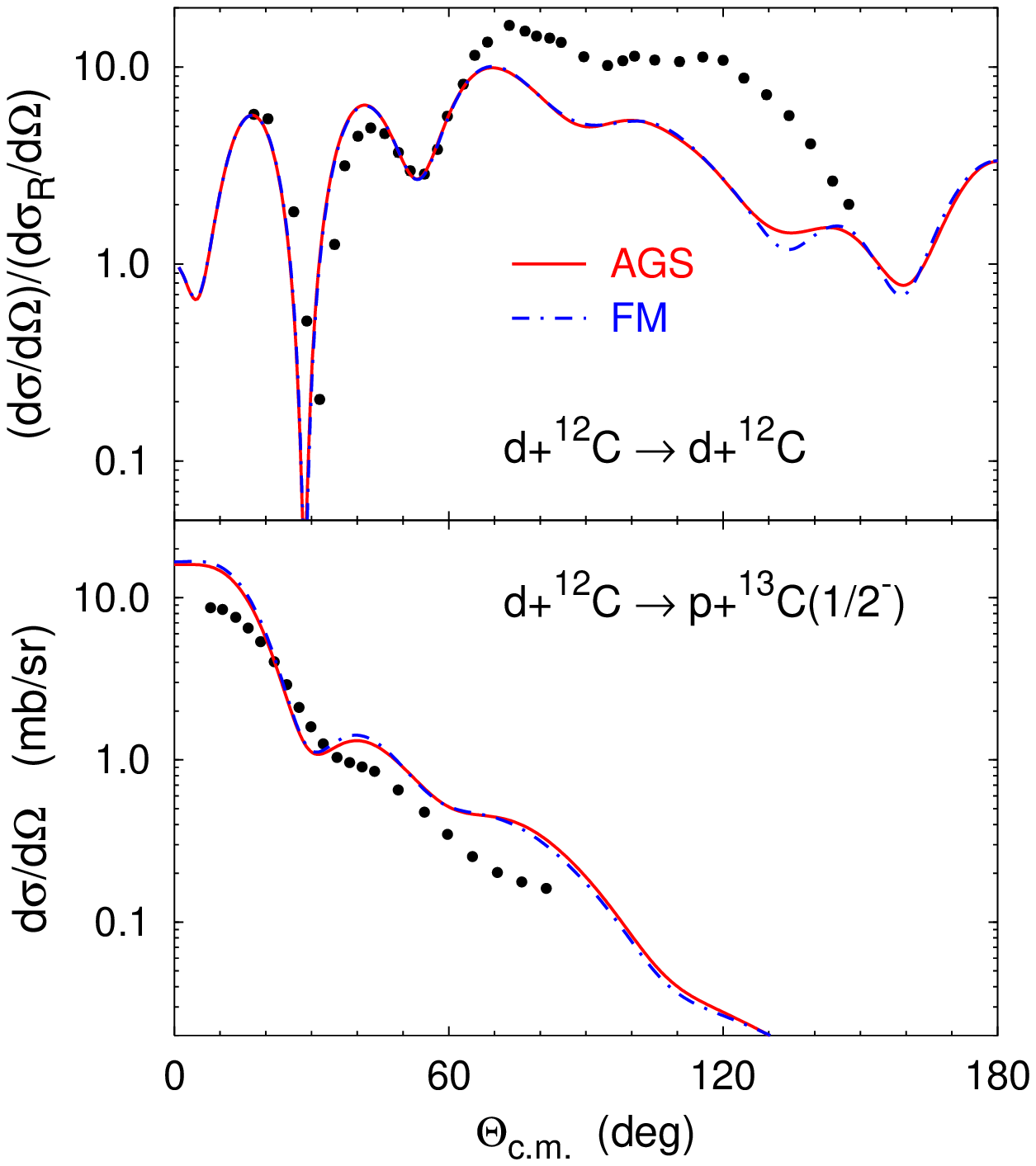}
\includegraphics[scale=0.47]{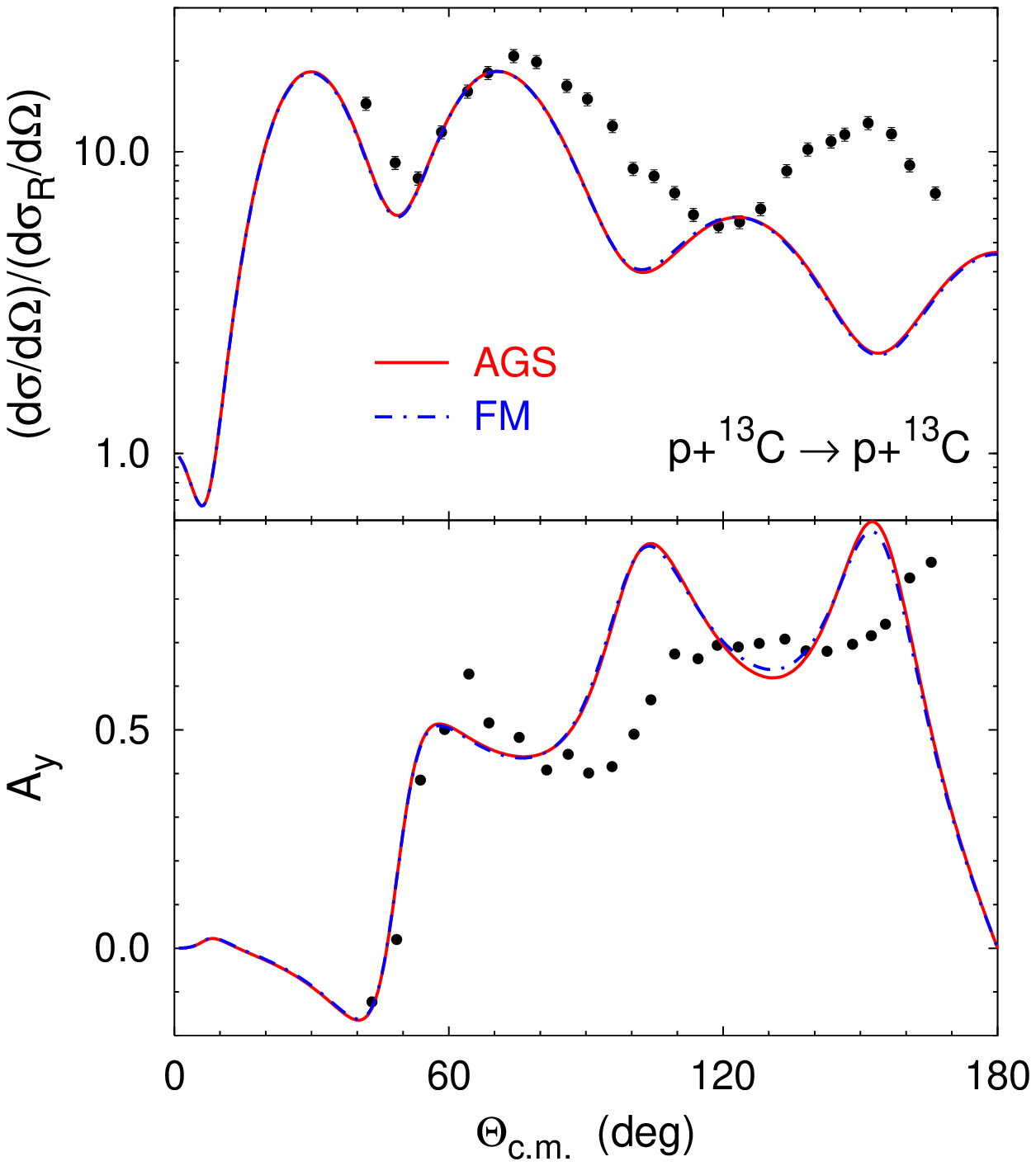}
\caption{\label{fig:dC} Comparison of the momentum space AGS results (solid curves)  and the current paper ones
(dashed-dotted curves). They  contain the
deuteron-${}^{12}$C scattering at 30 MeV deuteron laboratory energy (left panel)
and  the proton-${}^{13}$C elastic scattering at 30.6 MeV proton laboratory energy (right panel).
Differential cross sections for the elastic scattering and neutron stripping reaction are presented
 as a ratio to the Rutherford cross section $d\sigma_R/d\Omega$.
The proton analyzing power is displayed  in the bottom of the right panel figure.
The experimental data are  taken from Refs.~\cite{perrin:77,dC30p,pC30}.}
\end{center}
\end{figure}

More recently, calculations in 3-body sector have been extended to the study  of reactions including compound nucleus.
Here we present the results concerning ${d+^{12}C}\rightarrow p+^{13}C$ scattering, which were obtained
within a three-body model containing $n,p$ and $^{12}C$ clusters. The interaction between the
free nucleons is described by the realistic AV18 potential~\cite{AV18}, whereas the  $N-^{12}C$ interaction is described by
the optical CH89 potential~\cite{CH89}.
The $n$-$^{12}$C potential in the $^2P_\frac{1}{2}$ partial wave is made real
and is adjusted to support the ground state of $^{13}C$ with 4.946 MeV of  binding energy; the parameters are taken from
Ref.~\cite{nunes:11b}.
This adjustment is made in order to reproduce $p-{}^{13}$C threshold in the transfer reactions.
These results are summarized in Fig.~\ref{fig:dC} and compared  to
the ones obtained by the Lisboa group~\cite{Deltuva_AGS}, who used the formalism of
Alt-Grassberger-Sandhas (AGS) equations in momentum space.
One may see an excellent agreement between these two calculations for $d+ ^{12}C$ and $p+^{13}C$
elastic cross sections, as well as for the transfer $d+ ^{12}C\rightarrow p+^{13}C$  cross section.

\begin{figure}[!]
\begin{center}
\hspace{-0.5cm} \includegraphics[scale=0.57]{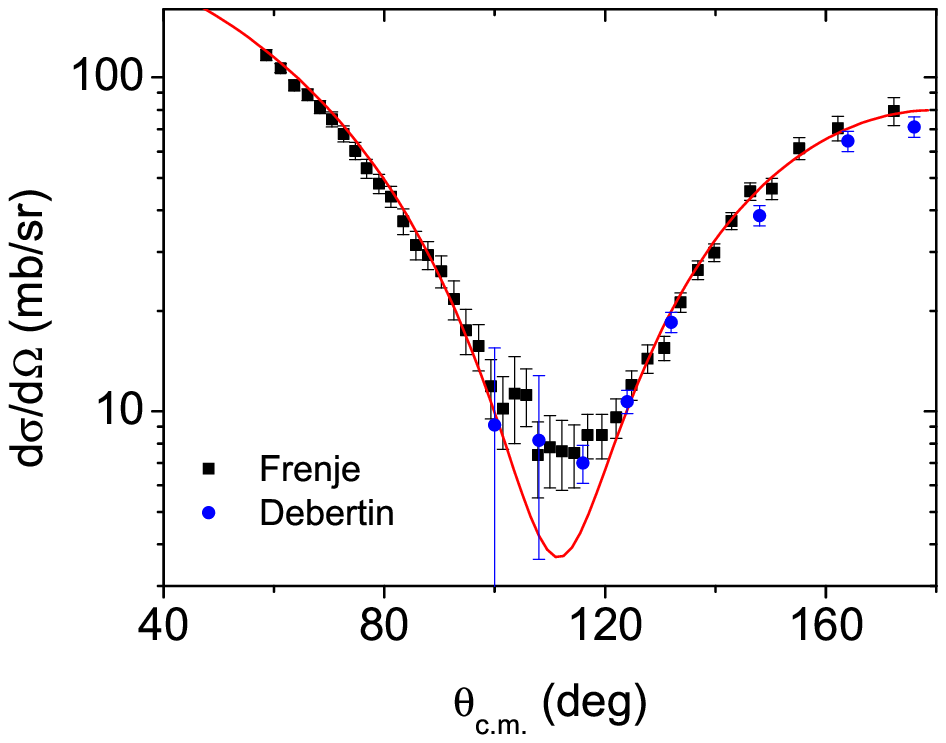}
\hspace{-0.5cm} \includegraphics[scale=0.57]{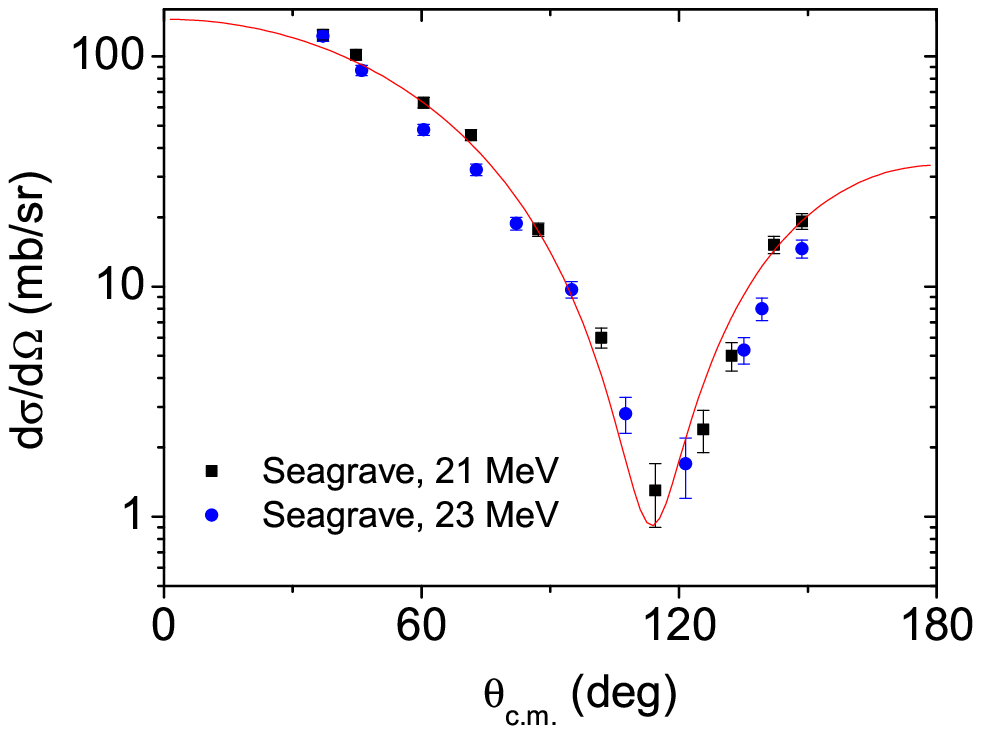}
\end{center}
\caption{Calculated $n-^3H$ elastic differential cross sections for neutrons of laboratory energy 14.4 MeV (left panel) and 22.1 MeV (right panel) compared with experimental results of Frenje et al.~\cite{Frenje_nt}, Debretin et al.~\cite{Debertin} and Seagrave et al.~\cite{Seagrave}.}
\label{fig:r_dep}
\end{figure}

Concerning the four-body case, we present in table~\ref{tab_4nt_cs} a calculation of the
integrated elastic and break-up cross sections for $n-^3H$ collisions. These results
are obtained for the total isospin $T=1$ channel, using MT I-III potential~\cite{MT13}.
A rather reasonable agreement is obtained with the experimental values, although
the agreement worsens at higher energies. MT I-III potential, being restricted to the
S-waves, is not appropriate to this high energy domain.
In figure~\ref{fig:r_dep}  the corresponding differential elastic cross sections are displayed,
calculated for incident neutrons at laboratory  energy 14.4 MeV (left panel) and 22.1 MeV (right panel).
One may notice that a rather good agreement is obtained also in this case. Only
at the minimum region, for 14.4 MeV neutrons, the theoretical results underestimate the experimental values.
These small discrepancies are due to the simplicity of the MT I-III potential.  It has
been shown recently~\cite{Arnas_cem:2012} that the realistic interactions
further improve the results, providing almost perfect agreement with  the data,
both for the differential as well as for the integrated elastic cross sections.

\begin{table}[tbp]
\caption{Neutron-triton elastic ($\sigma_e$), inelastic ($\sigma_b$) and total ($\sigma_t$) scattering cross sections (in mb)
for the selected neutron laboratory energies (in MeV) compared with the experimental data.}
\label{tab_4nt_cs}%
\begin{tabular}{cccccc}
\hline\noalign{\smallskip}
 $E_{lab}$ &   \multicolumn{3}{c}{MT I-III} &   \multicolumn{2}{c}{ Exp. } \\
  (MeV)  & $\sigma_e$ & $\sigma_b$ & $\sigma_t$ & $\sigma_t$ & [Ref.]  \\
   \noalign{\smallskip}\hline\noalign{\smallskip}
14.4           & 922    & 11  & 933 &978$\pm$70 &\cite{Battat}  \\
18.0           & 690    & 25  & 715 &750$\pm$40 &\cite{Battat}  \\
22.1           & 512    & 38  & 550 &620$\pm$24 &\cite{Phillips}\\
\noalign{\smallskip}\hline
\end{tabular}
\end{table}

 \section{Conclusions}

 A  method based on the complex scaling is presented to solve the scattering problem above the break-up thresholds.
 This method does not require the implementation of  highly non trivial boundary conditions and allows to solve the
 few-body scattering problem
 using square-integrable functions. Scattering problem might thus be solved using configuration space
 bound state techniques in complex arithmetics.

Reliable and accurate results have already been obtained for the 3- and 4-body elastic and break-up processes,
including optical potentials as well as long-range Coulomb repulsion.
This method opens a way to explore the many-body scattering and breakup  reactions and allows an accurate treatment
of a rich variety of problems in molecular as well as in nuclear physics.

 \begin{acknowledgements}
This work was granted access to the HPC resources of IDRIS under the allocation 2009-i2009056006
made by GENCI (Grand Equipement National de Calcul Intensif). We thank the staff members of the IDRIS for their constant help.
\end{acknowledgements}

\end{document}